# Indications of superconductivity in doped highly oriented pyrolytic graphite

**Grover Larkins and Yuriy Vlasov**
Department of Electrical and Computer Engineering, Florida International University, Miami, Florida 33174, USA

E-mail: larkinsg@fiu.edu

**Abstract.** We have observed possible superconductivity using standard resistance vs. temperature techniques in phosphorous ion implanted Highly Oriented Pyrolytic Graphite. The onset appears to be above 100 K and quenching by an applied magnetic field has been observed. The four initial boron implanted samples showed no signs of becoming superconductive whereas all four initial and eight subsequent samples that were implanted with phosphorous showed at least some sign of the existence of small amounts of the possibly superconducting phases. The observed onset temperature is dependent on both the number of electron donors present and the amount of damage done to the graphene sub-layers in the Highly Oriented Pyrolytic Graphite samples. As a result the data appears to suggest that the potential for far higher onset temperatures in un-damaged doped graphite exists.

## 1. Introduction

Carbon has large number of allotropes exhibiting different mechanical, chemical and electrical properties [1]. Recently many researchers have turned their attention to graphene [2], a two dimensional carbon structure that could be used as a test object to study possible electron pairing necessary for superconductivity. Highly Oriented Pyrolytic Graphite (HOPG) can be used in this experimental study since interlayer coupling in it is weak [3,4] and it approximates graphene fairly well. Since HOPG is a robust, easily handled material as well we decided to perform our initial investigation using it as a starting material.

## 2. Background

It has been hypothesized by us, and others [5-7], that the close coupling or strong scattering of electrons by both phonons and plasmons in graphene indicates a potential for superconductivity at considerable temperatures in doped graphene [8,9]. Kopelevich *et al.* [10,11] reported a few cases of suspected superconductivity in Highly Oriented Pyrolytic Graphite but, given that the results were only on as-grown samples, the origins and likelihood of the observations being due to superconductivity were unclear. Our work is the first that we are aware of where a systematic attempt has been made to substitutionally dope HOPG/graphene/graphite into a superconductive state. The work described herein represents our first results in an attempt to either confirm or disprove this hypothesis.

The difficulties involved with doping graphene and concerns with how to physically handle graphene in a testing environment motivated our concentration on Highly Oriented Pyrolytic Graphite. HOPG is easily handled and is structurally "stacked" graphene sheets. Two potential dopants, one an electron donor (phosphorous), the other an electron acceptor (boron) were selected. The simple expediency of ion implantation at low energies and doses was selected as a doping method in this simple set of experiments. The energy and doping levels were purposely selected to minimize the damage done by the implantation to the graphene sheets in the HOPG. This was done to reduce the disorder (damage) in the lattice as it was clear that this disorder could provide scattering centers that would very likely have a negative impact on any electron-electron coupling mechanism, regardless of whether that coupling mechanism is phonon or plasmon mediated. In short, it was recognized that any reduction in the electron-electron coupling mechanism would reduce the critical temperature or critical field of the material and would be best avoided if possible.

### 3. Experimental

The as-purchased grade ZYH MikroMasch and NT-MDT HOPG was first measured in a Janis/CTI closed cycle refrigeration system with an ultimate temperature of less than 15 K and our well tested four-point probe resistance vs. temperature ($R$ vs. $T$) system was connected to it. The actual measurements were taken using a Hewlett-Packard 4263A LCR meter running under computer control using our custom written LabView GPIB data collection program to control the entire run. After this testing the samples were then implanted with phosphorus or boron.

The implantation of the boron and phosphorous was performed by Cutting Edge Ions, LLC on a mail-in basis. The implantation energies and doses are shown in table 1. The computed depth profile of ion implanted phosphorous in graphite and the computed damage profile are shown in figure 1. Since there was no characterized implantation model for the stopping power of HOPG as a substrate we selected the nearest and most similar substrate material, graphite, for the simulation. Clearly this is somewhat less than ideal, however, for the simple task of estimating the range, damage and ion distribution in the HOPG the results should be accurate to within a few percent and that is sufficient for our purposes.

**Table 1.** A list of HOPG samples used in this work. All samples were 10 mm × 10 mm × 1.2 mm in size.

| HOPG sample number | Implant | Energy (KeV) | Implantation dose (cm$^{-2}$) | Substrate manufacturer |
|---|---|---|---|---|
| 001 | boron | 10 | $4.0\times10^8$ | MikroMasch |
| 002 | boron | 10 | $8.0\times10^8$ | MikroMasch |
| 003 | boron | 10 | $2.0\times10^9$ | MikroMasch |
| 004 | boron | 10 | $4.0\times10^9$ | MikroMasch |
| 005 | phosphorus | 10 | $1.2\times10^8$ | MikroMasch |
| 006 | phosphorus | 10 | $3.0\times10^8$ | MikroMasch |
| 007 | phosphorus | 10 | $5.9\times10^8$ | MikroMasch |
| 008 | phosphorus | 10 | $1.2\times10^9$ | MikroMasch |
| 009 | phosphorus | 10 | $1.2\times10^8$ | MikroMasch |
| 010 | phosphorus | 10 | $6\times10^7$ | MikroMasch |
| 011 | phosphorus | 10 | $1.2\times10^8$ | NT-MDT |
| 012 | phosphorus | 10 | $6\times10^7$ | NT-MDT |
| 013 | phosphorus | 5 | $1.2\times10^8$ | NT-MDT |
| 014 | phosphorus | 5 | $6\times10^7$ | NT-MDT |

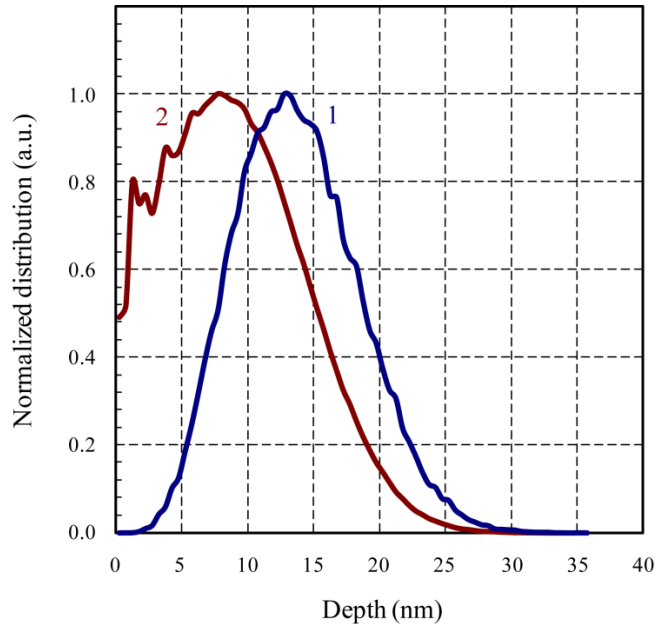

**Figure 1.** SRIM simulated distribution of (1) phosphorus ions implanted in graphite at $E_p$ = 10 KeV and (2) damage in graphite lattice caused by implant. (SRIM-2008 software developed by J.F. Ziegler is available for download at web site www.srim.org).

After implantation the sample was placed into the Janis/CTI closed cycle refrigeration system again and the $R$ vs. $T$ characteristic was re-measured.

In the event that a $R$ vs. $T$ signal that suggested the possible presence of a superconducting phase was observed we would use a permanent magnet with a surface field of approximately 0.5 T (0.3 T to 1 T) to either partially or totally quench the $R$ vs. $T$ signal, thus confirming superconductivity. The magnet was placed directly on the HOPG sample and held in place with a special fixture to keep it from migrating under the vibration of the refrigeration system.

## 4. Results and Discussion

Boron doped HOPG samples show no signs of possible superconductivity at any temperature down to below 20 K (two representative curves are shown in figure 2).

On the other hand, samples 005 and 011 (figure 3 and 4), the lowest dose phosphorous implanted HOPG samples, exhibit a deviation from the expected monotonic rise in resistance as temperature goes down at some point above 100 K (we are comfortable with that noting that the deviation actually begins above 120 K in several of the measurements we have taken). This and the fact that there is a fairly steep drop in the resistance (by a factor of more than 2) at lower temperatures was enough to be considered an indicator of possible superconductivity in the sample.

The magnet was then added to the system and the $R$ vs. $T$ characteristic of both samples 005 and 011 was then re-measured and the effect was observed to quench to a semiconductive type $R$ vs. $T$ characteristic with negative slope at low temperatures. This is shown in figures 3 and 4.

Both samples 005 and 011 were sent to Dr. Richard Greene's group at the University of Maryland and the characteristics were confirmed, this is shown in figure 5. It was hoped that the more sophisticated instrumentation available to Dr. Greene's group would be able to see the magnetic susceptibility signal characteristic of a superconductive transition, however, given the diamagnetism native to HOPG and the possibly extremely thin nature of the active layer this proved to be beyond the capability of their instrumentation.

A more detailed look at the other phosphorous doped samples led to the discovery that all twelve of them showed some sort of deviation towards lower resistance at low temperature but that none of them were as significant as that seen in samples 005 and 011. This, coupled with the fact that the entire effect could be quenched by a modest magnetic field, has led us to contemplate the possibility that the implantation damage may be sufficient to reduce the electron-electron coupling and that in samples 005 and 011 the lower dose and consequently lower damage level has allowed the coupling to overcome the effects of the damage. This chain of logic leads to the hypothesis that the observed effect is taking place in the "tail" of the implanted phosphorous distribution where there is less damage (figure 1, curve 1). Given that this layer is very likely only a few nanometers in thickness it is unsurprising that a modest magnetic field could quench the system (figure 5).

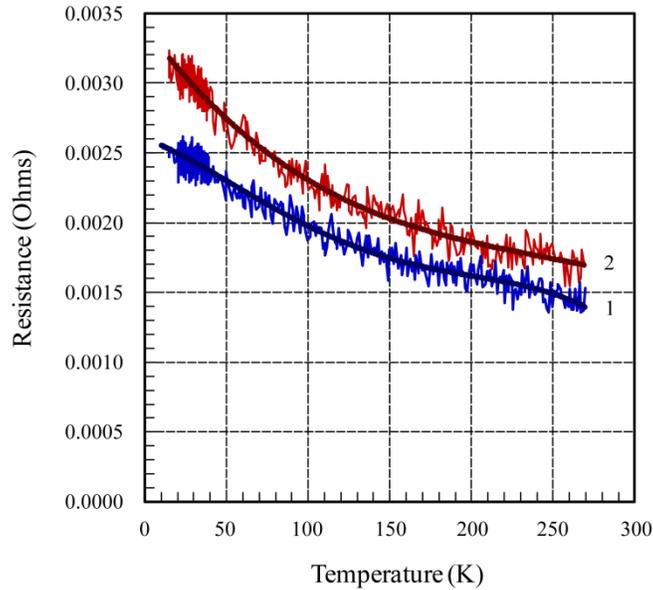

**Figure 2.** Measured $R$ vs. $T$ dependence of representative boron-doped HOPG samples. (1) sample 001. (2) sample 002. Smooth lines are a fourth order polynomial fit.

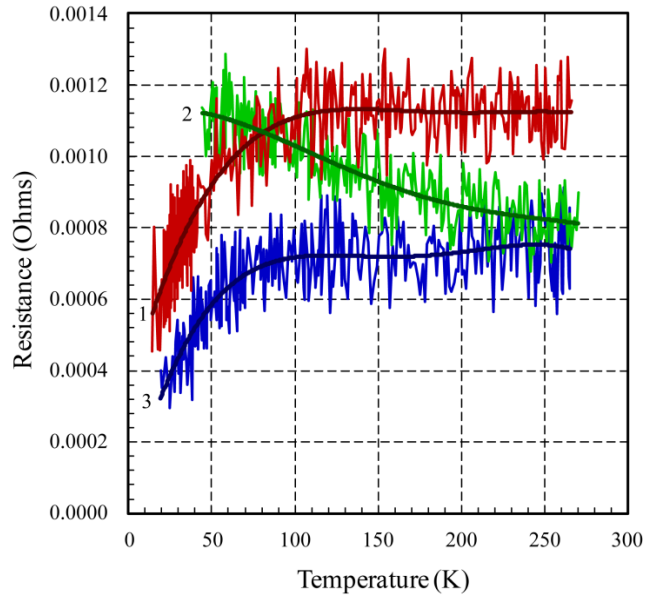

**Figure 3.** Measured *R* vs. *T* dependence of Phosphorus-doped HOPG sample 005. (1) before magnetic field was applied. (2) with magnetic field applied. (3) after magnetic field was removed. Smooth lines are a fourth order polynomial fit.

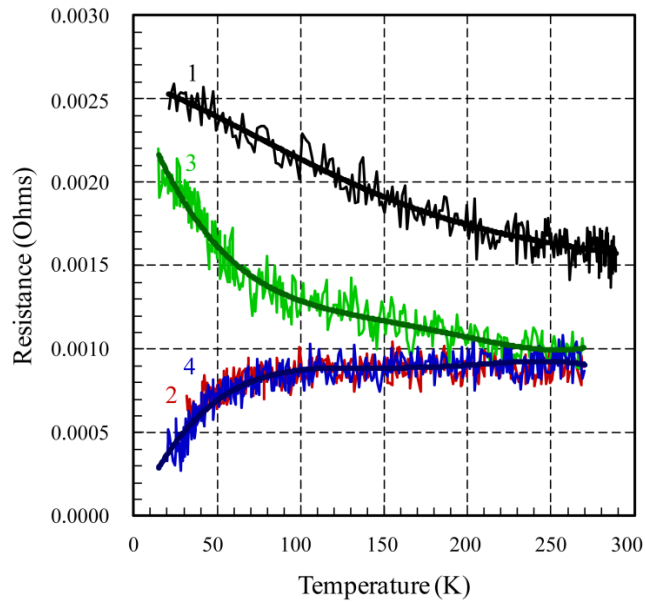

**Figure 4.** Measured *R* vs. *T* dependence of Phosphorus-doped HOPG samples 011. (1) before doping. (2) doped sample before magnetic field was applied. (3) with magnetic field applied. (4) after magnetic field was removed. Smooth lines represent corresponding polynomial fit.

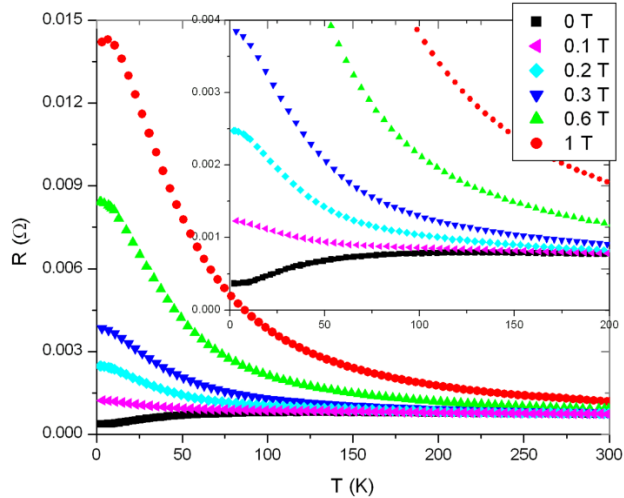

**Figure 5.** *R* vs. *T* measurement of phosphorus doped HOPG samples 005 in the presence of magnetic field from 0 T to 1 T. (These tests were performed at University of Maryland by Dr. Paul Bach).

## 5. Conclusion

We have observed a response suggestive of superconductivity in phosphorous (electron donor) implanted Highly Oriented Pyrolytic Graphite. The ultimate critical temperature in this system is in excess of 100 K and, may very likely be considerably higher if damage incurred during the doping can be further minimized. The observed "superconductive type" effect is very likely confined to a very thin layer somewhat further into the HOPG than the peak of the implantation distribution (figure 1). Doping with electron acceptors (boron) has not been observed to induce the effect despite their probably having caused less damage (lower mass, lower dose and same energy) to the HOPG than the phosphorus.

Both of these conclusions track well with theoretical hypotheses that electron donor doped graphene may become superconductive but, clearly more work needs to be done before the actual mechanism of the observed effect becomes completely clear. The results thus far however are of such a nature that we felt compelled to share them and we welcome further investigation into the matter by our colleagues.

**Acknowledgments**

The authors wish to thank Dr. Richard Greene and Dr. Paul Bach at University of Maryland who tested our samples to confirm the observed Resistance vs. Temperature curves which are shown in this work. This work was supported by the United States Air Force under AFOSR grant FA9550-10-1-0134.